\documentclass[pra,twocolumn,superscriptaddress,letterpaper]{revtex4}

\usepackage{amsmath,amsfonts}
\usepackage{graphicx}
\usepackage{enumerate}
\usepackage{hyperref}

\newcommand{\e}{\mathrm{e}}
\newcommand{\bh}{\mathbf{h}}
\newcommand{\bI}{\mathbf{I}}
\newcommand{\bS}{\mathbf{S}}
\newcommand{\be}{\mathbf{e}}
\newcommand{\mean}[1]{\langle #1 \rangle}
\newcommand{\bmean}[1]{\bigl\langle #1 \bigr\rangle}

\begin{document}


\title{
Nuclear Magnetism and Electronic Order in $^{13}$C Nanotubes
}

\author{Bernd Braunecker}
\affiliation{Department of Physics, University of Basel, 
             Klingelbergstrasse 82, 4056 Basel, Switzerland}

\author{Pascal Simon}
\affiliation{Department of Physics, University of Basel, 
             Klingelbergstrasse 82, 4056 Basel, Switzerland}
\affiliation{Laboratoire de Physique et Mod\'{e}lisation des  Milieux
             Condens\'{e}s, CNRS and Universit\'{e} Joseph Fourier, BP 166, 
             38042 Grenoble, France}
\affiliation{Laboratoire de Physique des Solides, CNRS UMR-8502,
             Universit\'{e} Paris Sud, 91405 Orsay Cedex, France}

\author{Daniel Loss}
\affiliation{Department of Physics, University of Basel, 
             Klingelbergstrasse 82, 4056 Basel, Switzerland}

\date{\today}



\begin{abstract}
Single wall carbon nanotubes grown entirely from $^{13}$C form an ideal system 
to study the effect of electron interaction on nuclear magnetism in one dimension. 
If the electrons are in the metallic, Luttinger liquid regime, we show that even 
a very weak hyperfine coupling to the $^{13}$C nuclear spins has a striking effect: 
The system is driven into an ordered phase, which combines electron and nuclear 
degrees of freedom, and which persists up into the millikelvin range. In this phase 
the conductance is reduced by a universal factor of 2, allowing for detection by 
standard transport experiments.
\end{abstract}


\maketitle

\begin{figure*}
	\includegraphics[width=1.5\columnwidth]{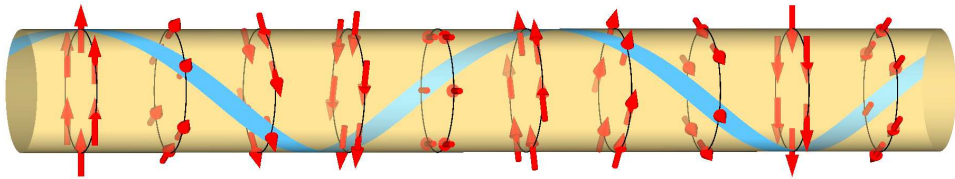}
	\caption{
	Illustration of the helical nuclear magnetism (indicated by the blue ribbon)
	of the single wall $^{13}$C nanotube (SWNT), which emerges below a critical 
	temperature through strong renormalization of the hyperfine coupling by electron
	correlations.
	The nuclear spins (red arrows) order ferromagnetically on a cross-section 
	of the SWNT and rotate along the SWNT axis with a period $\pi/k_F$ in the
	spin $xy$-plane (chosen here arbitrarily orthogonal to the SWNT axis).
	Through the feedback of the nuclear magnetization the electric conductance
	of the SWNT is reduced by a factor of precisely 2.
	\label{fig:tube}
	}
\end{figure*}

The physics of conduction electrons interacting with localized magnetic moments is central
for numerous fields in condensed matter such as nuclear magnetism \cite{froehlich:1940},
heavy fermions \cite{tsunetsugu:1997}, or ferromagnetic semiconductors \cite{ohno:1992,ohno:1998,dietl:1997,konig:2000}.
Nuclear spins embedded in metals offer 
an ideal platform to study the interplay between strong electron correlations 
and magnetism of localized moments in the RKKY regime.
In two dimensions the magnetic properties of the localized moments \cite{simon:2007,simon:2008} 
depend indeed crucially
on electron-electron interactions \cite{belitz:1997,hirashima:1998,maslov:2003,maslov:2006,shekhter:2006}. 
In one-dimensional (1D) systems such as single wall carbon nanotubes (SWNTs)
electron correlations are even more important. For metallic (armchair) SWNT they 
lead to Luttinger liquid physics \cite{egger:1997,kane:1997,bockrath:1999}.
Recently, SWNTs made of $^{13}$C, forming a nuclear spin lattice, have become experimentally 
available \cite{marcus:2008a,marcus:2008b,simonf:2005,ruemmeli:2007}. 
Motivated by this we study here nuclear magnetism in metallic $^{13}$C SWNTs. 
We show that even a weak hyperfine interaction can lead to
a helical magnetic order of the nuclear spins (see Fig. \ref{fig:tube}) coexisting with
an electron density order that combines
charge and spin degrees of freedom.
The ordered phases stabilize each other, and the critical temperature
undergoes a dramatic renormalization up into the millikelvin range
due to electron-electron interactions.
In this new phase the electron spin susceptibility becomes anisotropic
and the conductance of the SWNT drops by a universal factor of 2.

The drastic restructuring of the electron wave functions
through the renormalization is very different from the case
of two \cite{simon:2007,simon:2008} or three dimensions
\cite{froehlich:1940} where it is, in comparison, weak.
The same renormalization leads to considerable anisotropy in the 
electron system: The nuclear magnetic field spontaneously breaks
the spin rotational symmetry; it rotates in a plane, which we can 
associate with the spin $(x,y)$ directions (see Fig. \ref{fig:tube}). 
This plane is singled out as an easy-plane through the stabilization 
of the electron density wave, and electron correlation functions 
become anisotropic between the spin $(x,y)$ plane and the 
spin $z$ direction. We 
illustrate this behavior below
through the calculation of the electron spin 
susceptibilities.
We emphasize that this anisotropy is a crucial feature of
the SWNT system studied here and appears \emph{spontaneously}
due to strong renormalization of the RKKY interactions. This
distinguishes our system, in particular, from models with built-in
easy-axis anisotropy \cite{zachar:1996}.


\emph{Model.} ---
We assume that the electrons are confined in a single mode $\psi_\perp$ in the directions
perpendicular to the tube axis. 
The nuclear spins $I = 1/2$ of the $^{13}$C ions on a circular cross section 
have identical overlaps with this transverse mode, and so identical couplings
to the electrons. Through their indirect RKKY interaction over the electron
gas they are therefore locked in a ferromagnetic alignment (see Fig. \ref{fig:tube}). 
This RKKY interaction, described below,
overrules furthermore the direct dipolar interaction between the nuclear spins.
The latter is very small \cite{paget:1977}, $\sim 10^{-11}$ eV, and shall be neglected henceforth.
This allows us to treat the nuclear spins as a 1D chain 
of \emph{large} $\tilde{I} = I N_\perp $ spins, composed of the sum of 
the $N_\perp \sim 50$ spins around a circular cross section. 
Due to this, Kondo physics, which requires small quantum spins,
can be excluded from the beginning.

Hence, we model the SWNT by a 1D 
nuclear spin lattice of length $L$ coupled through the hyperfine
interaction to a 1D
electron gas. The Hamiltonian resembles that of a Kondo 
lattice $H = H_{el} + A \sum_i \hat{\bS}_i \cdot \hat{\bI}_i$,
where $i$ runs over the 1D lattice sites with positions $r_i$,
$\hat{\bI}_i = (\hat{I}_i^x,\hat{I}_i^y,\hat{I}_i^z)$ is the effective nuclear spin 
of size $\tilde{I} = I N_\perp$,
$\hat{\bS}_i = (\hat{S}_i^x,\hat{S}_i^y,\hat{S}_i^z)$ is the electron spin operator
at site $i$, and $A = A_0/N_\perp$ is the on-site hyperfine interaction
constant $A_0$ weighted by the transverse electron mode.
In contrast to the usual Kondo lattice model,
$H_{el}$ describes the \emph{interacting} electrons and is 
defined in Eq. \eqref{eq:H_el} below. 

The precise value of $A_0$ in SWNTs is unknown.
Estimates in the literature \cite{pennington:1996} provide values of 
$A_0 \sim 10^{-7}-10^{-6}$ eV, depending much on the curvature
of the nanotube 
(higher values have been reported in \cite{marcus:2008a} though).
This compares with the typical energy scales of the electrons,
which can be quantified by the value
$E_F = v_F k_F /2$ (we set $\hbar =1$ throughout this paper), 
where
$k_F /\pi = n_{el}$ is the electron density in the system and
$v_F$ ($\approx 8 \times 10^5$ m/s in SWNTs \cite{egger:1997,kane:1997,saito:1998}) 
is the typical velocity of electron excitations. 
Through the dependence on $n_{el}$, $E_F$ can vary between the meV 
to eV range.


\emph{Effective model.} ---
Due to the small ratio $A/E_F$, the energy and time scales 
related to the electrons and nuclear spins decouple, and we 
can treat both subsystems separately. 
A Schrieffer-Wolff transformation of $H$ allows us
to obtain an effective Hamiltonian for the nuclear spins
\cite{simon:2007,simon:2008},
\begin{equation}
	H^{eff}_n = \frac{1}{2} \sum_{ij\alpha} 
	\frac{J^\alpha_{ij}}{N_\perp^2} \hat{I}_i^\alpha \hat{I}_j^\alpha
	= \frac{1}{L}\sum_{q \alpha}\frac{J^\alpha_q}{N_\perp^2} \hat{I}^\alpha_{-q}\hat{I}^\alpha_{q},
\end{equation}
where $\alpha = x,y,z$, and
$J^\alpha_{ij} = A_0^2 \chi^{\alpha\alpha}_{ij} a / 2$ is the effective
RKKY \cite{kittel:1987} 
interaction between nuclear spins. $a$ is the lattice spacing and 
provides the short distance cutoff of the continuum theory.
The sum over $q = n \pi / L$ for integer $n$ runs over the first Brillouin zone.
$\chi^{\alpha\alpha}_{ij} = -i a^{-1} \int_0^\infty dt \ \mean{[\hat{S}_i^\alpha(t),\hat{S}_j^\alpha(0)]} \e^{-\eta t}$
(for an infinitesimal $\eta>0$) is
the static electron spin susceptibility. 
We also have defined
$\hat{I}^\alpha_q = \sum_i \e^{i r_i q} \hat{I}^\alpha_i$ 
and $J^\alpha_q = \int dr \ \e^{-i r q} J^\alpha(r)$.

The effective electron Hamiltonian, on the other hand, includes the effect of the feedback
of the nuclear field on the electrons. Since the spins $\tilde{I} = I N_\perp$
are large, we can choose 
$H_{el}^{eff} = H_{el} + H_{Ov}$, with 
$H_{Ov} = \sum_i \bh_i \cdot \hat{\bS}_i$ and
$\bh_i = A \mean{\hat{\bI}_i}$ the nuclear Overhauser field.


\emph{Interacting electrons as Luttinger liquid.} ---
We use a bosonized Hamiltonian to describe the interacting electron system
of the armchair SWNT, which is naturally in the Luttinger liquid state due 
to the linear electron dispersion \cite{egger:1997,kane:1997}. 
The unit cell of a graphite sheet contains two carbon atoms, which results
into a two-band description of the bosonized system. Since mixing between the bands
is essentially absent
\cite{egger:1997,kane:1997} 
we shall, however, focus on 
a single band only in order to avoid a heavy notation.
The bosonized single-band Hamiltonian reads 
\cite{egger:1997,kane:1997,giamarchi:2004}
\begin{equation} \label{eq:H_el}
	H_{el} =  
	\sum_{\nu=c,s} \int \frac{dr}{2\pi} 
	\left[ 
		\frac{v_\nu}{K_\nu} (\nabla \phi_\nu(r))^2  
		+ 
		v_\nu K_\nu (\nabla \theta_\nu(r))^2 
	\right],
\end{equation} 
where $\phi_{c,s}$ are boson fields such that
$-\nabla \phi_{c,s}\sqrt{2}/\pi$ express charge and spin density 
fluctuations, respectively.
$\theta_{c,s}$ are such that $\nabla \theta_{c,s}/\pi$ are 
canonical conjugate to $\phi_{c,s}$.
$v_{c,s} = v_F / K_{c,s}$ are charge and spin wave velocities, and 
$K_{c,s}$ are the dimensionless Luttinger liquid parameters.
For SWNTs \cite{egger:1997,kane:1997}, $K_c \approx 0.2$ .
If the electron spin SU(2) 
symmetry is maintained, $K_s = 1$, otherwise $K_s \neq 1$.

\begin{figure}
	\includegraphics[width=\columnwidth]{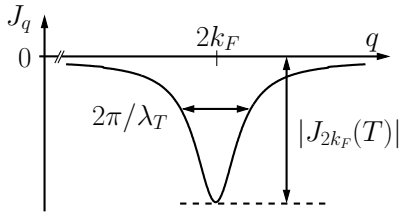}
	\caption{
	Sketch of the RKKY interaction $J_q$ given by Eq. \eqref{eq:J_q}.
	\label{fig:sketch}
	}
\end{figure}

\emph{Without feedback from nuclear magnetic field.} ---
Let us first assume that there is no feedback from the Overhauser field 
on the electrons and set $\bh_i \equiv 0$. The electron system 
forms a Luttinger liquid, for which the zero temperature 
spin susceptibility has a singularity at momentum $q = \pm 2k_F$
induced by backscattering processes \cite{egger:1996,giamarchi:2004}.
At $T>0$ this singularity turns into a steep but finite minimum:
The backscattering part of the 
spin operator $\hat{S}^x_i$ is expressed in the bosonization language
by the operators \cite{giamarchi:2004}
$\hat{O}_{SDW}^x(r_i) \propto \e^{-2ik_F r_i}\e^{i \sqrt{2}\phi_c}
\cos(\sqrt{2} \theta_s)$, such that
$\hat{S}^x = [\hat{O}_{SDW}^x + \hat{O}_{SDW}^{x\dagger}]/2$
plus forward scattering terms.
Similar expressions \cite{giamarchi:2004} hold for $\hat{S}^{y}$ and $\hat{S}^{z}$.
We further assume that $J^\alpha_q \equiv J_q$ is isotropic 
and in particular $K_s = 1$.
The correlators between those operators can be evaluated in the 
standard way and we obtain (for $q>0$)
\begin{equation} \label{eq:J_q}
	J_q(g,v_F) \approx - C(g,v_F) (k_B T )^{2g-2}
	\left| \Gamma(\kappa) / \Gamma(\kappa + 1-g)\right|^2,
\end{equation}
where 
$g = (K_c+K_s^{-1})/2$, 
$\kappa = g/2-i \lambda_T (q-2k_F)/4\pi$, depending on the
thermal length $\lambda_T = v_F / k_B T$ with
$k_B$ the Boltzmann constant. $\Gamma$ is Euler's Gamma function
and $C(g,v_F) = A_0^2 a \sin(\pi g) \Gamma^2(1-g) (2\pi a / v_F)^{2g-2} / 8\pi^2 v_F$.
We have made the inessential assumption $v_c = v_s = v_F$. 
Note that $J_q$ is independent of $k_F$ for a linear dispersion.
A density dependence of $J_q$ requires a curvature of the 
electron dispersion, which partially restores Fermi liquid 
properties \cite{khodas:2007}, a scenario which we 
disregard for metallic SWNTs.
A sketch of $J_q$ is shown in Fig. \ref{fig:sketch}.

At temperatures $T < T^*_0$ [defined in Eq. \eqref{eq:T*_0} below],
$|J_{2k_F}(T)| > k_B T$ and the nuclear
spins can -- classically -- minimize the RKKY energy by aligning in a spiral order
$\bI_i' = I N_\perp [ \cos(2k_F r_i) \be_x + \sin(2k_F r_i) \be_y]$, 
where $\be_{x,y}$ are orthonormal vectors defining the spin $(x,y)$ plane.
We shall henceforth \emph{assume} that this order is established, and 
show that this assumption is self-consistent. 
Fluctuations  reduce this maximal polarization, and in general
$|\mean{\hat{\bI}_i}| < I N_\perp$. 
The lowest lying excitations (to order $1/I N_\perp$) in the nuclear spin system
are magnons. 
Since $J_{ij}$ is long-ranged the energy cost of local 
defects, like kinks, scales with the system size and is very high.

For a helimagnet, there exists a gapless magnon band
with the dispersion \cite{simon:2008}
$\omega_q = 2 (IN_\perp) (J_{2k_F+q}/N_\perp^2 - J_{2k_F}/N_\perp^2)$.
Let 
$m_i = \mean{\hat{\bI}_i} \cdot \bI_i'/(IN_\perp)^2$
measure
the component of the average magnetization along $\bI_i'$,
normalized to $0 \le m_i \le 1$. 
Its Fourier component $m_{2k_F}$ acts as an order parameter for the spiral phase. 
Magnons decrease this order parameter and 
we have \cite{simon:2008}
\begin{equation} \label{eq:m_2kF}
	m_{2k_F}(T) = 1 - 
	\frac{a}{(I N_\perp)L} \sum_{q \neq 0} \frac{1}{\e^{\omega_q/k_B T} -1},
\end{equation}
where the sum represents the magnon occupation number.
In the continuum limit $L \to \infty$ the integrand is divergent as $1/q^2$
for $q \to 0$ (the $q=0$ mode is absent because the system is not a ring), 
showing the absence of true long range order in the 1D 
system.
Despite its appearance 
the divergence is not a consequence of the 
Mermin-Wagner theorem \cite{mermin:1966,bruno:2001}, which forbids long 
range order in low-dimensional systems for
sufficiently short ranged interactions. Since $J_{ij}$ is long ranged
this theorem does not apply.

\begin{figure*}
	\includegraphics[width=1.5\columnwidth]{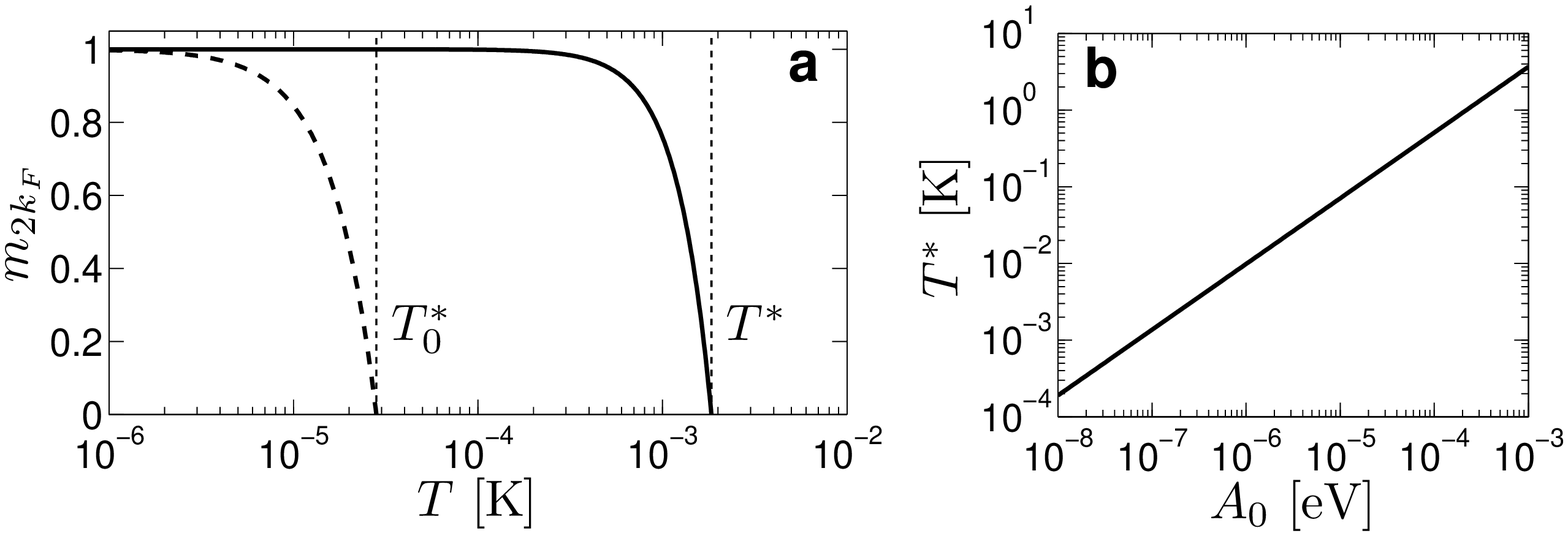}
	\caption{
	\textbf{a}: Magnetization $m_{2k_F}(T)$ [Eq. \eqref{eq:m_2kF_approx}]. 
	Dashed line: without feedback. Solid line: with feedback.
	Parameters for the curves are \cite{pennington:1996,egger:1997,kane:1997,saito:1998}
	$E_F=0.1$ eV, $A_0=10^{-7}$ eV, $v_F = 8 \times 10^5$ m/s,
	$a = 2.46$ \AA, $K_s = 1$, $K_c = 0.2$ 
	(leading to $g=0.6$, $g'=0.33$), 
	and $L=2$ $\mu$m. 
	The vertical lines mark the temperatures written next to them. 	
	\textbf{b}: Characteristic temperature $T^*$ [Eq. \eqref{eq:T*}] as a function 
	of the hyperfine constant $A_0$. The curve follows a power law 
	$T^* \propto A_0^{2/(3-2g')} = A_0^{0.86}$, and is plotted up to the self-consistency
	limit $T^* \approx v_F/L k_B = 3$ K.
	\label{fig:magn}}
\end{figure*}

The present situation, however, is very different in that the system has a 
finite length $L \sim 2$ $\mu$m imposed either
through the natural length of the nanotube or through an external confining 
potential.
At temperatures $T < T^*_0$ we find that $L \ll \lambda_T$, and 
so the cost of exciting the first magnon is already very high
$\omega_{q=\pi/L} \approx 2 I |J_{2k_F}(T)|/ N_\perp$.
We can define a temperature $T_{M0}$ providing the scale 
of the excitation of the first magnons by imposing
$\omega_q / k_B T \approx 2 I |J_{2k_F}(T)|/ N_\perp k_B T = 1$.
For $T > T_{M0}$ we can then simplify Eq. \eqref{eq:m_2kF} to 
\begin{equation} \label{eq:m_2kF_approx}
	m_{2k_F}(T) 
	\approx
	1 - \frac{1/IN_\perp}{\e^{(\frac{T_{M0}}{T})^{3-2g}}-1}
	\approx 
	1 - \left(\frac{T}{T^*_0}\right)^{3-2g},
\end{equation}
where we have defined
\begin{equation}  \label{eq:T*_0}
	k_B T^*_0 
	= 
	\left[ 2 I^2 C(g,v_F) 
	\Gamma^2(g/2) \Gamma^{-2}(1-g/2) \right]^{\frac{1}{3-2g}}.
\end{equation}
For the SWNT this temperature satisfies the self-consistency 
condition $k_B T_{M0} < k_B T^*_0 \ll v_F/L$.
We use $T^*_0$ as an estimate for the critical temperature. 
For a typical SWNT $T^*_0$ is very low.
With the values given with Fig. \ref{fig:magn} we obtain
$T^*_0 \sim 10$ $\mu$K, too low for experimental detection. 
Yet this analysis completely neglects the feedback of the magnetic
field on the electron gas. 
This leads to a strong renormalization of $T^*_0$.


\emph{Feedback of nuclear magnetic field on electrons.} ---
The ordering of the nuclear spins leads to a spatially oscillating
Overhauser field $\bh_i = A \mean{\bI_i}$
that acts back on the electrons. 
We choose the electron spin axis such that $\hat{\bS} \cdot \be_x = \hat{S}^x$
and $\hat{\bS} \cdot \be_y = \hat{S}^y$.
The spatial oscillations of $\bh_i \propto \e^{\pm 2i k_F r_i}$ in $H_{Ov}$
perfectly cancel some of the spatial oscillations of the $\hat{O}_{SDW}^{x,y}$
operators of the $\hat{S}_i^{x,y}$. 
Neglecting the remaining (irrelevant) oscillating terms 
we obtain 
$H_{Ov} \approx \sum_i A_0 I m_{2k_F} \cos(\sqrt{2 K}\phi_+(r_i))$,
where we have introduced
$\phi_+ = (\phi_c + \theta_s)/\sqrt{K}$
with the normalization $K= K_c + 1/K_s$.
The Hamiltonian becomes
of the sine-Gordon type and $H_{Ov}$ is relevant 
in the sense of the renormalization group (RG):
The $\phi_+$ field is pinned at a minimum of the cosine term of $H_{Ov}$.
The result is a density wave that combines 
charge and spin degrees of freedom.
Fluctuations about the minimum are massive, with a mass associated to
an energy scale $\Delta$. 
At commensurate electron filling Umklapp processes would
become relevant too, and lead to fully gapped charge and
spin sectors.
For SWNTs, however, this would
require high electron densities leading to $E_F \approx 1.4$ eV.
This case is not considered here.

Within a perturbative RG approach we find that 
\begin{equation}
	\Delta 
	\sim (A_0 I m_{2k_F}/E_F)^{1/(2-g)} v_F/a.
\end{equation}
This mass gap $\Delta$ is the first important consequence of the
feedback. The second important consequence is the spontaneous 
generation of anisotropy because the spin $(x,y)$ plane is singled
out by the Overhauser field.
This is seen, for instance, in the spin susceptibilities $\chi^{\alpha\alpha}$.
Those can be calculated in the same way as before (see Appendix A)
if we notice that the 
massive $\phi_+$ field does not contribute to the long-wavelength asymptotics.
The finite temperature expressions for the $\chi^{\alpha\alpha}$ 
are otherwise identical to the case without feedback, and the 
RKKY couplings $J_q^\alpha$ can be obtained from Eq. \eqref{eq:J_q}
upon the following modifications: 
For $\chi^{xx}$ and $\chi^{yy}$ the exponent $g$ is replaced by 
$g' = 2 K_c/K_s K$ and the amplitude is reduced by a factor 2 because 
a term depending on $\phi_+$ only drops out. 
For $\chi^{zz}$ the exponent becomes $g'' = (K_c/K_s + K_c K_s)/2K$
while the amplitude remains unchanged.
$v_F$ is replaced by $v_- = (v_c/K_c +v_sK_s)/K$.
This leads to
\begin{equation} \label{eq:J_q_ren}
	J_q^{x,y} = J_q(g',v_-)/2, \ 
	J_q^z = J_q(g'',v_-).
\end{equation}
For $K_c = 0.2$ and $K_s = 1$ we have to compare $g= 0.6$
with the strongly renormalized $g'=0.33$ and $g''=0.17$. 

Let us finally note that correlators between $\phi_+,\theta_+$ 
can only be neglected as long as $k_B T < \Delta$, 
i.e. $\lambda_T^{-1} < \xi^{-1}$ with $\xi = v_F/\Delta$
the correlation length.
In Eq. \eqref{eq:T*} below we define a critical temperature $T^*$ 
similarly to $T^*_0$ before.
For $T\ll T^*$,  $m_{2k_F} \approx 1$ (see Fig. \ref{fig:magn}), 
and we find that $\Delta \gg k_B T$. 
At $T \to T^*$, however, $m_{2k_F}$ vanishes and so does $\Delta$.  
The order in electron and nuclear systems, therefore, 
vanishes simultaneously.


\emph{Consequences for magnetization and transport.} ---
The helical order still minimizes the energy and there remains a gapless 
magnon band \cite{simon:2008}, $\omega_q = 2 I (J'_{2k_F+q}-J'_{2k_F})/N_\perp$,
where $J'_q = J^x_q = J^y_q$.
The previous discussion of the magnetization remains otherwise unchanged. 
Replacing $J_q$ by $J'_q$ in Eq. \eqref{eq:T*_0}  
leads to the renormalized critical temperature $T^*$,
\begin{equation} \label{eq:T*}
	k_B T^* 
	= 
	\left[ I^2 C(g',v_-) \Gamma^2(g'/2) \Gamma^{-2}(1-g'/2) \right]^{\frac{1}{3-2g'}}.
\end{equation}
The notable difference is the modified exponent. For the parameters
displayed with Fig. \ref{fig:magn}, we obtain the change
from $1/(3-2g) = 0.625$ to $1/(3-2g') \approx 0.43$. 
Quite remarkably this considerably 
boosts the value of the characteristic temperature from 
$T^*_0 \sim 10$ $\mu$K to 
$T^* \sim 1$ mK. Note that $T^* \ll v_F / L k_B$ is still satisfied.
Fig. \ref{fig:magn} (\textbf{a}, solid line) shows the result of the
feedback. 
In Fig. \ref{fig:magn} (\textbf{b}) we also show the dependence 
of $T^*$ on $A_0$, 
$T^* \propto A_0^{2/(3-2g')} =  A_0^{0.86}$.

The order furthermore modifies the transport properties of the system.
With the opening of the mass gap in the $\phi_+$ channel, half of 
the conducting modes are blocked and the conductance decreases  
by the universal factor of 2. As an illustration we consider a  
SWNT connected to metallic leads. The conductance is given by
\cite{maslov:1995,ponomarenko:1995,safi:1995}
$G = 4e^2/h$, where $e$ is the electron charge, $h$ the Planck constant,
and where 4 is the number of conducting channels (2 spin projections and 
2 bands). The pinning of the $\phi_+$ field
(in each band)
blocks 2 conductance channels and so reduces the conductance precisely by 
the factor 2 
(see Appendix B for details).
Such a reduction is a direct consequence of the nuclear 
spin ordering and the Luttinger liquid physics of the electrons,
and should be detectable experimentally in standard
transport setups.


As a conclusion, we emphasize that the physics described here
is quite general and is also relevant for other 1D
systems of the Kondo-lattice type.


This work was supported by the Swiss NSF and NCCR Nanoscience.
 

\appendix


\section{Susceptibilities in the partially gapped system}

The static susceptibilities $\chi^\alpha(q) = \chi^\alpha(q,\omega = 0)$
are the Fourier transforms of the
following retarded
electron spin response functions (we set $\hbar = 1$ throughout this section)
\begin{equation} \label{eq:chi}
	\chi^\alpha(r,t)
	= - i \Theta(t)
	\mean{[S^\alpha(r,t) \, , \, S^\alpha(0,0)]},
\end{equation}
where $S^\alpha(r,t)$ are the electron spin operators at position $r$
and time $t$, and $\Theta(t)$ is the step function.

In the bosonization language, the spin operators are expressed through the sum 
of forward scattering and backscattering operators as \cite{giamarchi:2004}
$S^\alpha = (S^\alpha)_{forw} + [O_{SDW}^\alpha + (O_{SDW}^\alpha)^\dagger]/2$. 
The forward scattering operators lead to a contribution to $\chi^\alpha$
that is regular in the momentum $q$, while the backscattering parts are 
singular or strongly peaked at momentum $q=2k_F$. We shall therefore 
neglect the forward scattering contribution and focus only on 
the backscattering operators (not writing Klein factors) \cite{giamarchi:2004} 
\begin{align}
	O_{SDW}^x 
	&= \frac{\e^{- 2ik_F r}}{2 \pi a} \e^{i \sqrt{2}\phi_c} 
	\left[ \e^{i\sqrt{2}\theta_s} + \e^{-i\sqrt{2}\theta_s} \right],
	\\
	O_{SDW}^y 
	&= i \frac{\e^{- 2ik_F r}}{2\pi a} \e^{i \sqrt{2}\phi_c} 
	\left[ \e^{i\sqrt{2}\theta_s} - \e^{-i\sqrt{2}\theta_s} \right],
	\\
	O_{SDW}^z 
	&= i\frac{\e^{- 2ik_F r}}{2\pi a} \e^{i \sqrt{2}\phi_c} 
	\left[ \e^{i\sqrt{2}\phi_s} - \e^{-i\sqrt{2}\phi_s} \right].
\end{align}
The average in Eq. \eqref{eq:chi} is evaluated with respect to the following Hamiltonian, depending on the 
gapped $\phi_+$ and ungapped $\phi_-$ fields,
\begin{equation} \label{eq:H_pm}
\begin{aligned}
	H 
	&= \int \frac{dr}{2\pi} 
	\Bigl\{ 
	v_- \Bigl[
		(\nabla \phi_-(r))^2 + (\nabla \theta_-(r))^2
	\Bigr]
 	\\
	&+ 
	v_+ \Bigl[
		(\nabla \phi_+(r))^2 + (\nabla \theta_+(r))^2
		+ \xi^{-2} \ \phi_+^2(r)
	\Bigr] 
	\Bigr\},
\end{aligned}
\end{equation}
where $\xi = v_+ / \Delta$ is the correlation length associated to the gap $\Delta$,
$v_+ = (v_c K_c + v_s/ K_s)/K$ and $v_- = (v_c/K_s + v_s K_c)/K$
for $K = K_c + 1/K_s$.

The boson fields $\phi_\pm, \theta_\pm$ 
are related to the usual charge and
spin boson fields through the transformation
\begin{align}
	\phi_c
	&= \frac{\sqrt{K_c}}{\sqrt{K}} 
	\left[ \sqrt{K_c} \phi_+ - \frac{1}{\sqrt{K_s}} \phi_- \right],
\label{eq:phi_c}
\\
	\phi_s
	&=  \frac{\sqrt{K_s}}{\sqrt{K}} 
	\left[  \frac{1}{\sqrt{K_s}} \theta_+ + \sqrt{K_c} \theta_- \right],
\label{eq:phi_s}
\\
	\theta_c
	&=  \frac{1}{\sqrt{K_c K}} 
	\left[   \sqrt{K_c} \theta_+ - \frac{1}{\sqrt{K_s}} \theta_- \right],
\label{eq:theta_c}
\\
	\theta_s
	&= \frac{1}{\sqrt{K_s K}} 
	\left[ \frac{1}{\sqrt{K_s}} \phi_+ + \sqrt{K_c}\phi_- \right].
\label{eq:theta_s}
\end{align}
Let us set $\bar{r} = (r,t)$, $\bar{\chi}^\alpha(\bar{r}) = -i 2 (a\pi)^2 \chi^\alpha(r,t)$,
and assume that $t>0$. We can then write, for instance for $\chi^x$,
\begin{align} 
	&\bar{\chi}^x(\bar{r})
	= 
		\cos(2k_F r) \Bigl[ 
		\bmean{\bigl[ \e^{i \sqrt{2K}\phi_+(\bar{r})} \, , \, \e^{-i \sqrt{2K}\phi_+(0)} \bigl]}
	\nonumber\\
	&\!+
		\bmean{\bigl[ \e^{i \sqrt{2K'} \phi_-(\bar{r}) - i \sqrt{2K''} \phi_+(\bar{r})}
		\, , \, \e^{-i \sqrt{2K'} \phi_-(0) + i \sqrt{2K''} \phi_+(0)} \bigl]}
	\Bigr],
\label{eq:chi_x_bar}
\end{align}
with $K' = 4 K_c/ K_s K$ and $K'' = (K_c-K_s^{-1})^2/K$, and where we have used the 
invariance of the Hamiltonian under a simultaneous sign change of all the boson fields.
In the Gaussian theory \eqref{eq:H_pm} the correlators in \eqref{eq:chi_x_bar} are fully expressed
through boson correlators of the form
$\mean{\phi_\pm(\bar{r}) \phi_\pm(0)}$ which, in $(q,\omega)$ space, 
are given by \cite{giamarchi:2004}
\begin{equation} \label{eq:phi_-_corr}
	\mean{\phi_-^*(q,\omega)\phi_-(q,\omega)}
	= 
	\frac{\pi v_-}{(\omega \pm i \eta)^2 - v_-^2 q^2},
\end{equation}
for the massless fields and
\begin{equation} \label{eq:phi_+_corr}
	\mean{\phi_+^*(q,\omega)\phi_+(q,\omega)}
	= 
	\frac{\pi v_+}{(\omega\pm i \eta)^2- v_+^2 q^2 - \Delta^2},
\end{equation}
for the massive fields. The sign the infinitesimal shift
$\pm i \eta$ is dictated by the time order of the operators.
Eq. \eqref{eq:phi_-_corr} is singular at $\omega \to 0,q \to 0$, and the 
proper treatment of this singular behavior leads to the 
power law behavior of the susceptibilities 
(at the shifted $q \to \pm 2k_F$) characteristic for a 
Luttinger liquid theory. 

On the other hand, Eq. \eqref{eq:phi_+_corr} is regular at $\omega \to 0,q \to 0$,
and so does not contribute to the power law divergence at $q \to \pm 2k_F$.
In fact, let us expand Eq. \eqref{eq:chi_x_bar} in powers of $\phi_+$.
The lowest nonzero term in the $\phi_+$ is 
\begin{align} \label{eq:corr_exp}
	&\cos(2k_F r) \Bigl[ 
		2 K \bmean{\bigl[ \phi_+(\bar{r}) \, , \, \phi_+(0) \bigl]}
	\nonumber\\
	&\!+
		2K'' \bmean{\bigl[ \e^{i \sqrt{2K'} \phi_-(\bar{r})} \phi_+(\bar{r})
		\, , \, \e^{-i \sqrt{2K'} \phi_-(0)} \phi_+(0) \bigl]}
	\Bigr].
\end{align}
The Fourier transform of the first term, depending on $\phi_+$ only,
at $\omega \to 0,q \to \pm 2k_F$ 
tends to a constant $\propto 1/\Delta^2$, and so contributes only 
insignificantly to the static susceptibility $\chi^x(q,\omega=0)$ 
at $q \approx 2k_F$. 
The second term involves a sum of products of the type
\begin{equation} \label{eq:corr}
	\cos(2k_F r)
	\e^{ K' \mean{\phi_-(\bar{r}) \phi_-(0)}}
	\mean{\phi_+(\bar{r}) \phi_+(0)}.
\end{equation}
Since $\phi_-$ is massless, the exponential factor evaluates
to a power law. With Eq. \eqref{eq:phi_+_corr} we can
then write the Fourier transform
of Eq. \eqref{eq:corr} as a convolution of the form
\begin{equation}
	\int dq' d\omega' 
	\left|\frac{1}{\omega'^2 - v_-^2 q'^2} \right|^{1-K'}
	\!\!\!
	\frac{1}{(\omega\!-\!\omega')^2 - v_+^2 (q_\pm\!-\!q')^2 - \Delta^2},
\end{equation}
where $q_\pm = q \pm 2k_F$.
For $\omega \to 0$ and $q_\pm \to 0$ we see that 
the $\omega'$ integral is dominated by the poles at $\pm \sqrt{q'^2+\Delta^2}$
and the weak singularities at $\pm q'$.
The contribution at the poles is more singular, and if we focus
on the pole at $\omega' = \sqrt{v_+^2q'^2+\Delta^2}$, 
we obtain
\begin{equation}
	\sim 
	\frac{1}{\Delta^{2(1-K')}} \int dq' \frac{1}{\sqrt{v_+^2q'^2+\Delta^2}}.
\end{equation}
The remaining integral leads to an arcsinh, which has an ultraviolet
divergence that has to be cut off at $1/a$. More importantly, however, 
the result has no infrared divergence, meaning that 
this expression remains regular at $q_\pm \to 0$. 

The latter results allow us to conclude that 
the Fourier transform of Eq. \eqref{eq:corr_exp}
is regular at $q \to \pm 2k_F$.
Since the theory \eqref{eq:H_pm} is Gaussian, higher order correlators are 
products of the latter results and so remain regular.
We have therefore shown that the singular behavior of the susceptibility
is fully controlled by the $\phi_+$ independent term in the expansion
of the $\e^{\pm i \phi_+}$, allowing us 
to use the approximation
\begin{align} 
	&\bar{\chi}^x(\bar{r})
	\approx 
		\cos(2k_F r)  
		\bmean{\bigl[ \e^{i \sqrt{2K'} \phi_-(\bar{r})}
		\, , \, \e^{-i \sqrt{2K'} \phi_-(0)} \bigl]},
\end{align}
which is of precisely the same form as the susceptibility of 
a regular Luttinger liquid.
The difference in the present case is the modified exponent
$K'$ and the fact that the first term in Eq. \eqref{eq:chi_x_bar}, 
depending on $\phi_+$ only, drops out. The amplitude 
of the resulting susceptibility $\chi^x(q)$ is reduced by a factor of 2.
With $g' = K'/2$ this leads to the zero temperature susceptibility \cite{giamarchi:2004}
\begin{equation}
	\chi^x(q) = - \frac{1}{2} \frac{\sin(\pi g')}{4 v_- \pi^2} \Gamma^2(1-g')
	\sum_\pm \left| \frac{2}{a (q \pm 2k_F)}\right|^{2-2g'}. 
\end{equation}
The susceptibility $\chi^y(q)$ involves the same combinations of the 
$\phi_+$ and $\phi_-$ fields and is identical to $\chi^x(q)$.
The susceptibility $\chi^z(q)$ is expressed through
the combinations 
$\phi_c \pm \phi_s 
= \frac{1}{\sqrt{K}} 
\left(
	K_c\phi_+ - \sqrt{\frac{K_c}{K_s}} \phi_- \mp \theta_+ \mp  \sqrt{K_c K_s} \theta_-  
\right)$.
In contrast to $\chi^{x}$ and $\chi^{y}$, the massless fields do not cancel
out for one of the $\pm$ signs, and so
the previous reduction of the amplitude by the factor 2 does not occur.
Again we can neglect the contributions from the massive $\phi_+$ and $\theta_+$ fields.
The resulting exponent of the power law in $\chi^z(q)$ then
depends only on the combination of the prefactors of the $\phi_-$ and $\theta_-$
fields, and is given by 
$g'' = (K_c/K_s + K_c K_s) / 2 K$ so that
\begin{equation}
	\chi^z(q) = - \frac{\sin(\pi g'')}{4 v_- \pi^2} \Gamma^2(1-g'')
	\sum_\pm \left| \frac{2}{a (q \pm 2k_F)}\right|^{2-2g''}. 
\end{equation}
The extension to temperatures $T>0$ is straightforward \cite{giamarchi:2004} and the
result is given by Eqs. \eqref{eq:J_q} and \eqref{eq:J_q_ren}. The main 
effect is that the singularity at $q=\pm 2k_F$ turns into a finite
minimum.

As temperature rises, the depth of this minimum decreases. The 
neglected $\phi_+$ correlators become important when the temperature becomes 
comparable to $\Delta$. 
For temperatures below $T^*$, the helical magnetization $m_{2k_F}$ is
close to 1 and so $\Delta \approx (A I/E_F)^{1/(2-g)} v_F/ a$.
For the values (see Fig. \ref{fig:magn}) $A/E_F = 10^{-6}, v_F = 8 \times 10^{5}$ m/s,
$a = 2.46$ \AA{} and $g = K/2 = 0.6$, we obtain
$\Delta \approx 6.8 \times 10^{-5}$ eV, i.e. a corresponding temperature
of $0.7$ K. This is much higher than the critical temperature $T^*$ and confirms
the validity of the approximations above.


\section{Reduction of the electrical conductance}

We illustrate here the reduction of the conductance of the 
nanotube by a universal factor 2 with the calculation of the 
electrical DC conductance $G$ of a Luttinger liquid that is connected to 
metallic leads. It was shown in  \cite{maslov:1995,ponomarenko:1995,safi:1995}
that in such a system the conductance is given by $G = \frac{e^2}{h} n$,
where $n$ is the number of conducting channels in the Luttinger liquid.
The Luttinger liquid properties appear in this formula only through the number
$n$ of channels, while the prefactor $e^2/h$ is entirely determined
by the properties of the leads.

A metallic single wall carbon nanotube has $n=4$, which is composed of a factor 2 arising from 
the two spin directions, and a factor 2 because the unit cell has two 
carbon atoms leading to two electron bands. 
With the feedback from the ordered nuclear magnetic field the $\phi_+$ 
field is pinned and we expect that it no longer contributes
to the electrical conduction. Since there is a $\phi_+$ in each 
band, the number of conducting channels then reduces to $n=2$.

In the following we shall prove this intuitive result:
As in \cite{maslov:1995,ponomarenko:1995,safi:1995} we model
the leads by one-dimensional Fermi liquids with $K_c = K_s = 1$.
The validity of the 
quadratic Hamiltonian \eqref{eq:H_pm} can then be extended into the leads
by introducing a spatial dependence on $v_\pm$ and $\xi$ (or $\Delta$) such that
$v_\pm(r) = v_F$ and $\xi(r) = \Delta(r) = 0$ when $r$ lies in the leads.

In the bosonization formulation the DC conductance can be 
calculated through the $\omega \to 0$ limit of the
nonlocal conductivity \cite{giamarchi:2004}
\begin{equation}
	\sigma(r,r';\omega) = -i \frac{4 e^2}{h} (\omega + i \eta) 
	g^r(r,r';\omega),
\end{equation}
where $g^r(r,r';\omega)$ is the Fourier transform of the retarded
boson Green's function
\begin{multline}
	g^r(r,r;t) 
	= - i \Theta(t) \mean{[ \phi_c(r,t) \, , \, \phi_c(r',0) ]}
\\	= L_+(r,r') g^r_+(r,r';t) 
	+ L_-(r,r')	g^r_-(r,r';t),
\end{multline}
with 
\begin{equation}
	g^r_\pm(r,r;t) 
	= - i \Theta(t) \mean{[ \phi_\pm(r,t) \, , \, \phi_\pm(r',0) ]}
\end{equation}
and
\begin{align}
	L_+(r,r') &= 
	\frac{K_c(r) K_c(r')}{\sqrt{K(r) K(r')}}, 
	\\
	L_-(r,r') &=
	\sqrt{\frac{K_c(r) K_c(r')}{K_s(r) K_s(r') K(r) K(r')}}.
\end{align} 
As found in \cite{maslov:1995,ponomarenko:1995,safi:1995}
the conductance depends only on the properties of the leads,
where $L_{\pm}(r,r') = 1/2$. If the system were entirely gapless,
both $\phi_-$ and $\phi_+$ channels would therefore contribute equally to the
total conductance, with $2e^2/h$ from each channel (the latter factor 2 being due to 
the two bands).

The gap $\Delta$ affects only the $\phi_+$ fields, and so 
the contribution $2 e^2/h$ from the two $\phi_-$ channels
remains unchanged. The contribution from the gapped $\phi_+$ fields, however,
drops to zero:
If we follow the method of Ref. \cite{maslov:1995}
the inclusion of the finite gap $\Delta(r)$ in the calculation of 
$g^r_+$ is straightforward. 
The result is
\begin{equation}
	g^r_+(r,r';\omega)
	= \frac{1}{2 \sqrt{\Delta^2-\omega^2}} \e^{-|r-r'|\sqrt{\Delta^2-\omega^2} / v_+},
\end{equation}
not writing further regular terms expressing reflections of propagating waves
on the ends of the nanotube, and 
for $r$ and $r'$ both lying in the nanotube.
In the DC limit $\omega \to 0$ we then see that the corresponding
conductance is zero,
\begin{equation}
	(\omega+i\eta) g^r_+(r,r';\omega) \to 0,
\end{equation}
while ungapped fields have here the nonzero limit $i/2$.
Hence the DC conductance is entirely determined by the two $\phi_-$ channels
from the two bands of the nanotube and is given by 
\begin{equation}
	G = 2 \frac{e^2}{h}.
\end{equation}


\end{document}